\documentclass[aps,floats,twocolumn,superscriptaddress,floatfix,nofootinbib]{revtex4}
\usepackage{amsfonts,amssymb,amsmath,bm}
\usepackage{graphics,epsfig}

\usepackage{here}
 \newcommand{\ue}{\mathrm{e}}
\newcommand{\ui}{\mathrm{i}\,}

 \def\un{\hbox{{1\kern -0.25em\raise
0.4ex\hbox{{\scriptsize $|$}}}}} 
\def\nset{\hbox{{I\kern -0.18em N}}}

\begin{document}

\newif\iffigs 
\figstrue
\iffigs \fi
\def\drawing #1 #2 #3 {
\begin{center}
\setlength{\unitlength}{1mm}
\begin{picture}(#1,#2)(0,0)
\put(0,0){\framebox(#1,#2){#3}}
\end{picture}
\end{center} }



\title{A Borel transform method for locating singularities of Taylor and
  Fourier series}
\author{W. Pauls}
\affiliation{CNRS UMR 6202, Observatoire de la C\^ote d'Azur, BP 4229, 06304
  Nice Cedex 4, France}
\affiliation{Universit\'e de Nice--Sophia--Antipolis, Nice, France}
\affiliation{Fakult\"at f\"ur Physik, Universit\"at Bielefeld,
Universit\"atsstra{\ss}e 25, 33615 Bielefeld, Germany}
\author{U. Frisch}
\affiliation{CNRS UMR 6202, Observatoire de la C\^ote d'Azur, BP 4229,
06304 Nice Cedex 4, France}
\affiliation{Universit\'e de Nice--Sophia--Antipolis, Nice, France}
\draft{J.\ Stat.\ Phys., in press} 
\date{\today}

\begin{abstract}
Given a Taylor series  with a finite radius of 
convergence, its Borel transform defines an entire function. A theorem of P\'olya relates the large distance 
behavior of the Borel transform in different directions to 
singularities of the original function. With the help of the new asymptotic
interpolation method of van der Hoeven, we show that from the knowledge
of a large number  of Taylor coefficients we can identify 
precisely the
location of such singularities, as well as their type when they are
isolated. There is no risk of getting artefacts with this method, which 
also gives us access to some of the singularities
beyond the convergence disk. 
The  method can also be applied to
Fourier series of analytic periodic functions and is here tested on
various instances constructed from solutions to the Burgers equation.
Large  precision on scaling exponents (up to twenty
accurate digits) can be achieved.
\end{abstract}
\vspace*{5mm}
\maketitle

\section{Introduction}
\label{s:introduction}

In the late nineteenth Century, 
Pincherle \cite{pincherle} and 
then Borel \cite{borel99,borel28}  introduced what
is now known as the Borel transformation: given a  formal 
series in powers of the complex variable $Z$
\begin{equation}
f(Z) = \sum_{n=0}^{\infty} a_nZ^n,
\label{deffz}ß
\end{equation}
one introduces the Borel transformed series
\begin{equation}
F(\zeta) \equiv \sum_{n=0}^{\infty} \frac{a_n}{n!} \zeta ^n\; .
\label{defFzeta}
\end{equation}
Since, for ${\rm Re}\, Z>0$,
\begin{equation}
\int _0^\infty \zeta ^n \ue ^{-Z\zeta}\, d\zeta = \frac{n!}{Z^{n+1}}\;,
\label{laplacepower}
\end{equation}
it is useful to introduce the function
\begin{equation}
f^{\rm BL} (Z) \equiv \frac{1}{Z} f\left(\frac{1}{Z}\right) = \sum_{n=0}^{\infty} \frac{a_n}{Z^{n+1}}\;,
\label{BLdef}
\end{equation}
which is formally the Laplace transform of $F(\zeta)$ and which in this
context is sometimes called the Borel--Laplace transform of $F$.

Borel's motivation was predominantly to give a meaning to divergent series 
such as  $\sum n! Z^n$ and  the Borel transformation has been extensively 
used to resum divergent series appearing in physics 
(see e.g.\ Refs.~\cite{LZ90,shawyerwatson,truc}).

In 1929 P\'olya \cite{polya29} observed that the  Borel transformation 
can also be used to obtain information about singularities of a Taylor
series (in powers of $1/Z$) with a finite radius of convergence, in which
case the function $F(\zeta)$ is entire. 
He proved a theorem relating
the convex hull of singularities of $f^{\rm BL}(Z)$ (the smallest convex set 
outside of which the function is analytic) to a function called the indicatrix
of $F(\zeta)$, roughly the rate of exponential growth  at infinity 
of $F(\zeta)$ as a function of the direction (for precise definitions
see Section~\ref{s:polya}). 


Here we show that this theorem can be used in conjunction with
high-accuracy numerical methods to obtain very precise information 
on singularities of Taylor and Fourier series. Singularities play an
important role in fluid dynamics and condensed matter physics
(see \cite{vandyke,guttmann,blue} and references therein). Using
P\'olya's theorem to devise a practical numerical method would not have been 
possible without recent progress in high-precision numerical
algorithms and, foremost,
the new technique for asymptotic interpolation of van der Hoeven 
\cite{jorasint} which can sometimes give  remarkable precision
(close to twenty digits) on scaling exponents.

The paper is organized as follows. In Section~\ref{s:darboux} we
recall some known facts about Taylor and Fourier series and their
singularities. In
Section~\ref{s:joris} we give a presentation of van der Hoeven's 
method from an applied mathematics point of view and show how it
works in practice, using known results for the Burgers equation. In Section~\ref{s:polya} we give
an elementary introduction to P\'olya's theorem.  
In Section~\ref{s:BPH} we present our new method, which we propose to
call BPH (Borel--P\'olya--Hoeven), for determining
the convex hull of singularities and, for the case of isolated
singularities on this hull, their positions and type. In  
Section~\ref{s:burgers-multimode} 
we test BPH using again the Burgers equation. 
In  Section~\ref{s:conclusion}
we 
discuss open problems and make concluding 
remarks.

\section{From Taylor and Fourier coefficients to singularities}
\label{s:darboux}

We first recall the close relation between Fourier and Taylor
series for analytic functions. Let $u(x)$ be a $2\pi$-periodic 
function which is analytic in some neighborhood of the real axis in
which it can be extended to a function $u(z)$, where $z=x+ \ui y$. After 
subtraction of a suitable constant we can assume that $\int_0^{2\pi}
u(x)dx =0$. The Fourier-series representation of $u$ reads
\begin{eqnarray}
u(x)&=& \sum_{k=\pm 1,\,\pm 2,\ldots} \ue ^{\ui kx} \hat u_k\;,
\label{Fx}\\
\hat u_k &=& \frac{1}{2\pi}\int_0^{2\pi} \ue ^{-\ui kx} u(x)dx\;.
\label{Fk}
\end{eqnarray}
We denote by $u ^+(x)$ (resp.\ $u ^-(x)$) the partial sum
of the Fourier series \eqref{Fx} with $k>0$ (resp.\ $k<0$), which
is analytic in the upper (resp.\ lower) half plane $y\ge 0$
(resp.\ $y \le 0$). Each of these two functions can be written as
a Taylor series by an exponential change of variable:
\begin{eqnarray}
u ^+(z)&=& \sum_{k>0} \hat u_k Z^k,\qquad Z\equiv \ue ^{\ui z}\;,
\label{tayloruplus}\\
u ^-(z)&=& \sum_{k>0} \hat u_{-k} \tilde Z^k,\qquad \tilde Z\equiv \ue ^{-\ui z}\;.
\label{tayloruminus}
\end{eqnarray}
Obtaining the singularities of an analytic periodic function from its 
Fourier coefficients is just basically the same problem as obtaining 
the singularities of an analytic function $f(Z) = \sum_{n=0}^{\infty}
a_nZ^n$ from its Taylor coefficients $a_n$. Hadamard's formula gives
us the radius of convergence of the Taylor series, namely the distance
to the origin of the nearest singularity(ies). If we happen to
know that this is an isolated singularity at $Z_\star$, we can relate the 
singular behavior  near $Z_\star$ to the asymptotic behavior of the
$a_n$ by the Darboux theorem \cite{darboux,henrici,hunterguerrieri}.
For this one assumes that, in a  neighborhood of $Z_\star$, the function
$f(Z)$  is given by
\begin{eqnarray}
f(Z) &=& \left(1-Z/Z_\star\right)^{-\nu}r(Z)+ a(Z), \quad \nu \ne 0,\,-1,\,
-2\,\ldots\;,\\
r(Z)&=& \sum_{k=0}^\infty b_k \left(1-Z/Z_\star\right)^k\;,
\end{eqnarray}
where the functions $r(Z)$ and $a(Z)$ are analytic in some disk
centered at the origin with a radius exceeding $|Z_\star|$. It then
follows that, for large $n$,
\begin{equation}
a_n \simeq \sum_{k=0}^\infty \frac{(-1)^k b_kZ_\star^{k-n}
  \Gamma(n+\nu-k)}
{n!\Gamma(\nu -k)}\;.
\label{darboux}
\end{equation}
The leading term is simply $a_n\simeq b_0
n^{\nu-1}/(Z_\star^n) \Gamma(\nu)$.  Applied to the Fourier
series \eqref{Fx}, the leading-order Darboux formula can be recast as
follows:  a branch-point
singularity with exponent $-\nu$ of 
$u(z)$ at a location $z_\star$ in the lower complex plane implies 
that for $k \to +\infty$  the Fourier coefficient $\hat u_k$ is 
asymptotically proportional to $k ^{\nu -1} \ue ^ {-\ui k z_\star}$.
This can be shown directly by applying standard  steepest
descent asymptotics to the integral \eqref{Fk}
\cite{carrier-krook-pearson}.

When the radius of convergence of a Taylor series is determined by
a single singularity of the type assumed by Darboux, the knowledge
of a sufficiently large number of Taylor coefficients with enough
accuracy permits an accurate determination of the position and type of the
singularity.  This can be done by an iterative algorithm  developed
by Hunter and Guerrieri \cite{hunterguerrieri} or by
the asymptotic interpolation method discussed in
Section~\ref{s:joris}.

Sometimes there are two Darboux-type singularities on the convergence
circle or, equivalently, the periodic function $u ^+(z)$ has two
singularities with the same imaginary part.  The interference of the 
two singularities produces then a sinusoidal modulation of the
Taylor coefficients. This can still be handled by an iterative 
algorithm  \cite{hunterguerrieri}, but not directly by the 
asymptotic interpolation method, for reasons explained in 
Section~\ref{ss:joris-comments}.
The BPH method of Section~\ref{s:BPH} can handle not only the case
of two or more isolated singularities on the convergence circle
but also ``hidden'' singularities located beyond this circle (or within this
circle if the series is in inverse powers of $Z$), whose contributions
to the Taylor coefficients are exponentially smaller than any term  in
\eqref{darboux}. From an asymptotic point of view these contribution 
are ``beyond all orders''.

\section{The asymptotic interpolation method}
\label{s:joris}

Suppose that we have a function  $G(r)$ of a scalar positive
variable $r$ 
for which we suspect that it has, for large $r$, an asymptotic 
expansion with a leading term  $C r^{-\alpha}\ue ^{-\delta r}$, as in
the Darboux theorem
\eqref{darboux}, but that we only
know its values numerically with high accuracy (tens to hundreds of
known
digits) on a regular grid
$r_0,\,2r_0,\ldots, N r_0$ with a large number $N$ of points (from fifty
to thousands, depending on the problem). We set
\begin{equation}
G_n \equiv G(nr_0),\quad n =1,2,\ldots,N\;.
\label{defGn}
\end{equation}
Can we determine parameters such as $C$, $\alpha$ and $\delta$ with
high accuracy? 
One way is of course just to ignore the subleading corrections
and to try a least square fit of the data to the functional form
$Cr^{-\alpha}\ue ^{-\delta r}$, after taking a logarithm. One then
  has the awkward problem of having to pick a fitting interval of
  values of $n$; the procedure usually gives poor accuracy and the
determination of subleading corrections is almost impossible.

A better way, used for example in Refs.~\cite{zinnjustin,shelley,paulsetal}, is to notice
that, if we take the second ratio, defined
as 
\begin{equation}
R_n \equiv \frac{G_{n}G_{n-2}}{G_{n-1}^2}= 
\left(1-\frac{1}{(n-1)^2}\right)^{-\alpha}\;,
\label{defRn}
\end{equation}
then both the constant $C$ and the exponential drop out. Assuming then
$n$ to be sufficiently large that we can ignore
subleading corrections, we  obtain
\begin{equation}
\alpha = -\frac{\ln R_n}{\ln (1-1/(n-1)^2)}\;.
\label{ratiologs}
\end{equation}
The other two parameters $C$ and $\delta$ appearing in  
\eqref{Fwithonesubleading} are then easily determined. If the
remainder, 
that is the discrepancy between the 
value of $\alpha$  predicted by \eqref{ratiologs}, which we denote
 $\alpha_n$, and its limit
$\alpha_\infty$ for $n\to \infty$, tends to zero in a known functional
way, e.g., exponentially or algebraically, then we
can  extrapolate the $\alpha_n$'s  to infinite values of
$n$ using, e.g., one of Wynn's algorithms \cite{wynnrho,wynnepsilon} (see,
Ref.~\cite{weniger} for a review of extrapolation methods). We shall come
back briefly to such issues in Section~\ref{ss:joris-comments}.
Without knowing something about the functional form
of the subleading corrections  which control the remainder, 
extrapolation may not work very well because the
choice of the appropriate algorithm depends on the functional form of
the remainder.

Recently, van der Hoeven introduced the \textit{asymptotic
interpolation} method \cite{jorasint} which allows in principle
the determination of the  asymptotic expansion of $G_n$ beyond 
leading-order terms. When the function $G_n$ is known with very high
precision and up to sufficiently large values of $n$, parameters such as the
scaling exponent $\alpha$ can sometimes be determined with extreme
accuracy, as we shall see in Section~\ref{ss:joris-burgers}. An important
feature of the asymptotic 
interpolation method is that it uses the determination of subleading terms
to improve the accuracy on leading-order terms. 

Here we shall just give a  short  elementary introduction to  the asymptotic
interpolation method for the case when the data $G_n$ are real
numbers. There are several variants of the asymptotic interpolation
method; ours differs occasionally from that of Ref.~\cite{jorasint}.
The basic idea of the asymptotic interpolation method is to perform
simple ``down'' transformations on the data $G_n$ which successively strip off
leading and subleading terms. After a number of such down steps 
which depends on the quality of the data, the transformed data
become sufficiently simple  to allow a straightforward interpolation step. 
The list of down transformations which are needed is given hereafter.
\begin{itemize}
\item[{\bf I}] Inverse:  $G_n \longrightarrow \frac{1}{G_n}$
\item[{\bf R}] Ratio: $G_n \longrightarrow \frac{G_n}{G_{n-1}}$
\item[{\bf SR}] Second ratio: $G_n \longrightarrow 
\frac{G_nG_{n-2}}{G_{n-1}^2}$
\item[{\bf D}] Difference: $G_n \longrightarrow G_n -G_{n-1}$
\end{itemize}
At each stage, tests are applied to decide which of the four transformations
should be applied in order to favor the stripping process as much as
possible.
If $|G_n| <1$ for large $n$, apply {\bf I}; otherwise
proceed. If $|G_n|$ grows ``slowly'' at large $n$ (we found that a useful
operational definition is to see if the growth can be
identified as algebraic with a rather well defined exponent),
apply {\bf D}; otherwise (``fast'' growth), apply {\bf R}. 
In addition, if  $|G_n|$ grows
or  decreases exponentially at large $n$, we found that it saves time
to apply {\bf SR}; also, if $|G_n|$ is a slowly decreasing function, it is more
convenient to apply $-${\bf D}.  Note that the differences or ratios
involved in {\bf R}, {\bf SR} and {\bf D} are backward; this conveniently keeps
the maximum index $N$ fixed.

When the procedure is iterated,
after a while, an ``interpolation stage'' is reached where the data can be
asymptotically interpolated in a simple fashion, typically by a constant plus
a small remainder tending to zero at large $n$. Basically this means that
we have successfully stripped off a certain number of terms in the asymptotic 
expansion.  For the kind of data which we
are considering here, the most useful interpolation stages usually arise
at  the sixth and thirteenth stages
(counting the original data as stage zero). 

There are two effects which 
limit the number of stages which can be applied to a given set of data.
First, whenever a ratio or a
difference are taken, the precision of the data (i.e.\ the relative rounding
error) deteriorates; as the number of transformations applied increases, 
rounding errors make the data increasingly noisy,
beginning usually with the highest values of $n$.  Second, the interpolation
stages require sufficiently large values of $N$,  since the constant
asymptotic behavior at large $n$ may be preceded by non-trivial transients.  
For a given resolution $N$ and a given precision,
the procedure must be stopped at the latest interpolation stage not
significantly affected by  the two effects just mentioned. In practice
we should have a significant range of values of $n$ 
over which the data are almost constant and not affected by rounding noise (if
the rounding noise is very low this range may extend all the way to $N$).   
When the down process is stopped the  data are interpolated and
the process is reversed, by applying ``up'' transformations which are the
inverses of the down transformation in the reverse order. The inverses of the
{\bf D}, {\bf R} and {\bf SR} transformations involve one or two unknown
additive or multiplicative constants which are determined using the highest
known values of the $G_n$ and of their down transforms. 

When the process is completed, the data are asymptotically expressed
as a truncated transseries.  Roughly, a transseries is a formal
asymptotic series involving integer or fractional powers, logarithms,
exponentials and combinations thereof \cite{ecalle,joristhesis,jorisspringer}.

A worked example will now give the reader a more concrete feeling.

\subsection{Testing the asymptotic interpolation method on the Burgers equation with a single-mode initial condition}
\label{ss:joris-burgers}

Here and in Section~\ref{s:burgers-multimode} we shall perform tests
using the  one-dimensional inviscid Burgers equation 
\begin{equation}
\partial_t u(t,x) +u(t,x)\partial_x u(t,x) =0\;,
\label{burgers}
\end{equation}
with a $2\pi$-periodic real initial condition $u_0(x)$ having a finite
number of Fourier harmonics. We begin by recalling some well-known
facts about the solution and the singularities of the inviscid Burgers
equation. Eq.~\eqref{burgers} has an
implicit solution in Lagrangian coordinates
\begin{equation}
u(t,x) = u_0(a); \quad x= a+tu_0(a)\;.
\label{implicit}
\end{equation}
Up to the time $t_\star$ of the appearance of the first shock, the Lagrangian
map $a \mapsto x$ has a Jacobian 
\begin{equation}
J(t,a)\equiv 1+t\partial_au_0(a)
\label{jacobian}
\end{equation}
which does not vanish in the real space domain and
\eqref{implicit} defines a unique real solution. This solution has 
singularities in the complex domain (with the real part defined modulo
$2\pi$) at locations which are the images by the Lagrangian map of the
zeros of the Jacobian $J$. Generically these are simple zeros. The 
singularities in Eulerian coordinates are then square-root branch points. 
The solution can also
be written explicitly using the Fourier--Lagrangian representation
\cite{FF83,BF01}, which in a special case was actually discovered earlier
by Platzman \cite{platzman}. In the periodic case, the simplest
representation, called the third Fourier--Lagrangian representation,
valid for $k\ne 0$, is
\begin{eqnarray}
u(t,x) &=& \sum_{k=-\infty}^\infty \ue ^{\ui kx} \hat u_k(t)\;,
\label{fourieru}\\
\hat u_k(t) &=& -\frac{1}{2 \ui \pi kt} \int_0^{2\pi}
\ue ^{-\ui k\left(a+t u_0(a)\right)} da\;.
\label{fl3}
\end{eqnarray}

In this section we take the ``single-mode'' initial condition
\begin{equation}
u_0(a)= -\frac{1}{2} \sin a\;,
\label{onehalfsin}
\end{equation}
for which the first real singularity is at $t_\star =2$. 
Using \eqref{fl3} and the integral
representation of the Bessel function $J_n$ of integer order $n$ 
(see, e.g., Ref.~\cite{AS65} p.~360), one finds \cite{platzman}
\begin{equation}
\label{Bessel2}
\hat{u}_k (t) = \frac{\ui}{ kt} J_k (kt/2)\;.
\end{equation}

For convenience, we shall consider the solution at $t=1$. This 
single-mode solution has
only one pair of complex conjugate singularities on the imaginary axis at
\begin{equation}
z_\star^\pm =  \pm \ui \delta,\qquad 
\delta = \ln \left(2+\sqrt 3\right) -\frac{\sqrt 3}{2}\;.
\label{zstaronemode}
\end{equation}

Bessel functions of large order and arguments have an asymptotic
expansion (in the sense of Poincar\'e), obtained through the method of steepest
descent by Debye  \cite{Debye}. (The matter is also discussed  in Chap.~VIII
of Ref.~\cite{watson-bessel}). Debye identified various asymptotic
regimes which, in our notation, depend on whether $t$ is less or larger
than $t_\star$ and is or is not very close to $t_\star$; his classification
is in one-to-one correspondence with that of the various
regimes relating to preshocks, as discussed, for example, in 
Refs.~\cite{FF83,BF01}.  When $t=1$, well before $t_\star$,  the
relevant Debye expansion for $k\to +\infty$ is:
\begin{equation}
\label{Debye1}
\hat u_k(1)  \simeq \frac{\ui}{\sqrt{\pi\sqrt 3 } }
k^{-\frac{3}{2} } \ue^{-\delta k}\left(
1 + \sum_{n=1}^{\infty } \frac{\gamma _n ( 2/\sqrt 3 ) }{k^n } \right) \;,
\end{equation}
where $\delta$  is given by \eqref{zstaronemode},
\begin{equation}.
\label{Debye2}
\begin{split}
& \gamma _1 (\xi ) = \frac{3\xi - 5\xi ^3}{24} , \\
& \gamma _2 (\xi ) = \frac{81 \xi ^2 - 462 \xi ^4 + 385 \xi ^6 }{1152} ,\\
& \gamma _3 (\xi ) = \frac{30375 \xi ^3 - 369603 \xi ^5 + 765765 \xi ^7 - 
425425 \xi ^9 }{414720} , \\
\end{split}
\end{equation}
and the higher-order polynomials $\gamma_n(\xi)$ satisfy recurrence 
relations given, e.g., in Ref.~\cite{AS65}. The leading term of this expansion
follows also from the Darboux theorem \eqref{darboux}.

Let us now show that the asymptotic interpolation method, as outlined in 
Section~\ref{s:joris}, when applied to the Fourier
coefficients of the single-mode solution \eqref{Bessel2} can recover
a suitably truncated version of the Debye expansion  \eqref{Debye1}. We  use
all the Fourier coefficients with $k= 1,\ldots,N$, where
$N=1000$ and define our initial data set as $G_n \equiv \hat
u_n(1)/\ui$.

Each coefficient is calculated with an 80-digit precision (using
Mathematica$^{\hbox{\scriptsize\textregistered}}$ and 120-digit
working precision). The basic transformations and their inverses are
implemented numerically in 80-digit precision, using the high-precision 
packages GMP  and MPFR available from \texttt{http://www.swox.com/gmp/}
and \texttt{http://www.mpfr.org/}\;.

\begin{figure*}
 \iffigs 
 \centerline{%
 \includegraphics[scale=1.35]{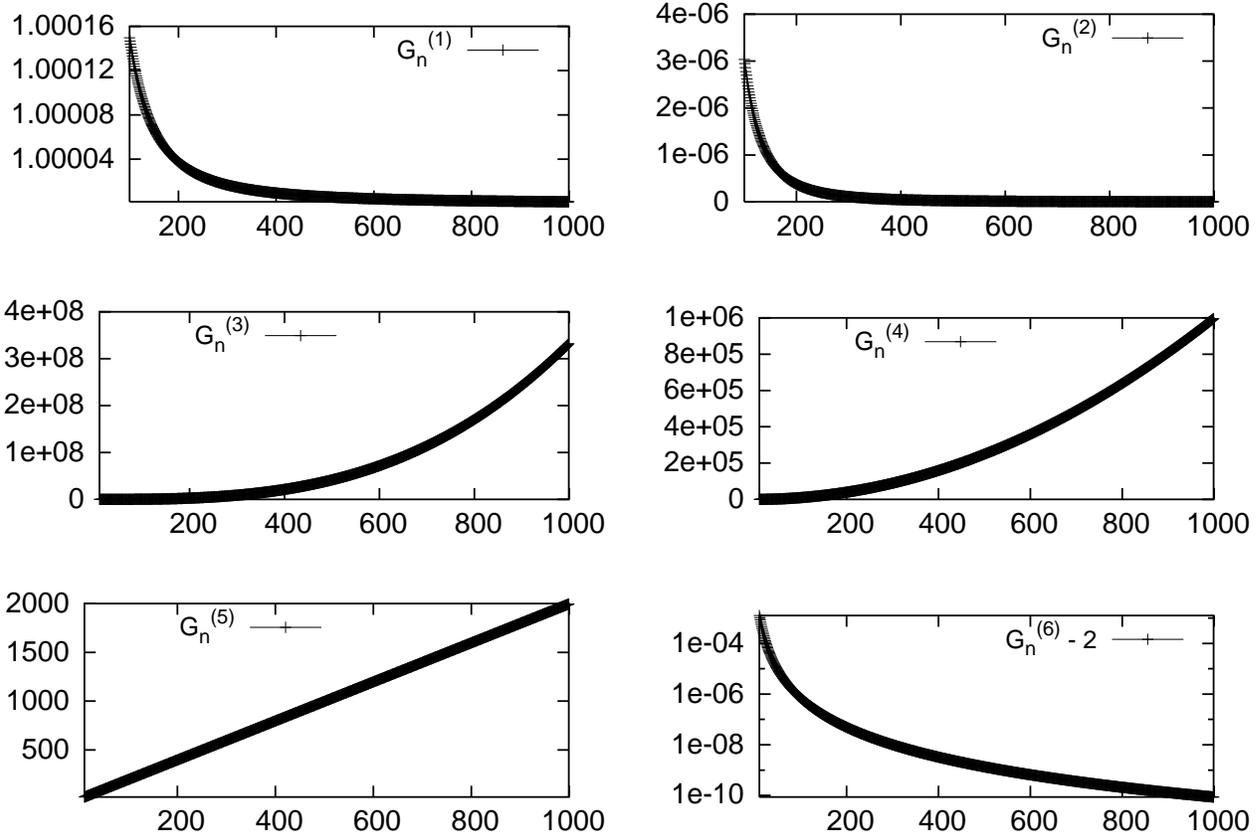}%
 }
 \else\drawing 65 10 {soc lin-log 2/1}
 \fi 
 
\caption{Numerical output from asymptotic interpolation at stages
$1$--$6$. Stages $1$-$5$ are represented in linear coordinates. For stage $6$
we represent the difference between $G_6 (n)$ and its asymptotic value $2$ in
lin-log coordinates.  }
\label{f:miniatures}
\end{figure*}

With these data we are able to reach stage 13. The list of
successively applied transformations, resulting from the tests given
in Section~\ref{s:joris}, is
\begin{equation}
\hbox{{\bf SR, -D, I, D, D, D, D, I, D, D, D, D, D}}
\label{listtransforms}
\end{equation}
Fig.~\ref{f:miniatures} shows the first six stages. It  is mostly
intended to bring out overall features and to make clear
which of the four transformations is to be selected at the next stage

It is very easy to
understand why the first six stages are as listed above. Indeed, let us
suppose that, to leading order, $G_n=C n^{-\alpha} \ue ^{-\delta
  n}$. We can work out analytically the various transforms and we list
hereafter the result up to the sixth stage, displaying only the
leading
and when needed the first subleading term in the large-$n$ expansion:
\begin{eqnarray}
&&\!\!\!\!C n^{-\alpha} \ue ^{-\delta   n}\; \stackrel{\bf SR}{\longrightarrow}\;
  1+\frac{\alpha}{n^2}\; \stackrel{\bf -D}{\longrightarrow}\;  \frac{2\alpha}{n^3}
\; \stackrel{\bf I}{\longrightarrow}\;  \frac{n^3}{2\alpha}\nonumber\\
&&\!\!\!\!\stackrel{\bf D}{\longrightarrow}\;  \frac{3n^2}{2\alpha}\;  \stackrel{\bf D}{\longrightarrow} \; 
\frac{3n}{\alpha}\; \stackrel{\bf D}{\longrightarrow}\;  \frac{3}{\alpha}
\label{sixfirst}
\end{eqnarray}
It is seen that stages 1 and 6 are interpolation stages
at which the data are asymptotically flat. Stage 6 is particularly
important since the asymptotic value $3/\alpha$ gives the
scaling exponent $\alpha$.  According to \eqref{Debye1}, for the Burgers
single-mode solution, the asymptotic value should be $2$.

Let us now show in some detail how the asymptotic interpolation technique
works to give us the asymptotic expansion of $G_n$. We begin  by limiting
ourselves to a six-stage procedure. The successively transformed data will
be denoted $G^{\rm (1)}$, \ldots, $G^{\rm (6)}$. 
Following Ref.~\cite{jorasint}, we interpolate $G^{\rm (6)}$ by $3/\alpha$.
How cleanly this can be done is visible in the last of the graphs 
in  Fig.~\ref{f:miniatures}, where we show the discrepancy  between 
$G_n^{\rm (6)}$ and its asymptotic value $2$. This discrepancy falls to
about $10^{-10}$ at the upper end of the range. Then we determine  
$G^{\rm (5)}$ by inverting the relation $G^{\rm (6)} = 
\hbox{\bf D}G^{\rm (5)}$. This involves an unknown additive constant which is
determined from the last data point $G_N^{\rm (5)}$. Then we continue inverting
the {\bf D} operators appearing at stages 4 and 5, each time using the 
last point to obtain
the additive constant. In this way we obtain a cubic polynomial for 
$G^{\rm (3)}$. We then invert the operator {\bf I}  and obtain the inverse
of the aforementioned cubic polynomial, which can be written
$-2\alpha n^{-3}(1 + d_1/n + d_2/n^2 +d_3/n^3+\ldots)$ with in principle
well defined constants $d_1$, $d_2$, etc. Then we invert the operator 
{\bf SR};  this can be done by taking a logarithm which will transform
second ratios into second increments. At the end of the process we obtain
the asymptotic expansion
\begin{eqnarray}
&&G_n\simeq C n^{-\alpha} \ue ^{-\delta n}\nonumber\\
&&\times \left[ 1 + \frac{\gamma _1 }{n} +
\frac{\gamma _2 }{n^2 } + \frac{\gamma _3 }{n^3 } + O\left( \frac{1}{n^4 }\right)\right]\;.
\label{asgzero}
\end{eqnarray}

It is actually simpler to start with \eqref{asgzero} and to apply
successively the first six transformations listed in \eqref{listtransforms}
to identify the parameters. With the six-stage procedure we obtain
$C$, $\alpha$, $\delta$, $\gamma_1$, $\gamma_2$,  $\gamma_3$. Their
values are given in Table~\ref{t:9coeff}. 
\begin{table*}
\hspace{-1.0cm}
\def\arraystretch{1.2}
\begin{tabular}{|c|l|l|l|} \hline
             & $\alpha $ & $\delta $ & $C$  \\ \hline
$6$ stages   & $ 1.49999999993 $ 
& $0.4509324931404 $  &
             $  0.4286913791     $  \\ \hline
$13$ stages  & $ 1.49999999999999995 $
& $0.450932493140378061868  $  & 
               $ 0.4286913790524959  $  \\ \hline
Theor.\ value & $3/2$
& $0.450932493140378061861  $            &
               $0.42869137905249585643 $        \\ \hline \hline
  & $\gamma _1 $ & $\gamma _2 $  &
$\gamma _3 $  \\ \hline
$6$ stages &
$ - 0.17641252$ &
$  0.17295$ &
$ -0.401 $ \\ \hline
$13$ stages & 
$-0.17641258225238$ & $0.172968106990 $ & $-0.406446182$ \\ \hline
Theor. value & $-0.176412582252385$ & $0.1729681069958$ &
$-0.4064461802$ \\ \hline \hline
  & $\gamma _4 $ & $\gamma _5 $  &
$\gamma _6 $  \\ \hline
$13$ stages & 
$1.384160933  $ & $- 6.192505762$ & $  34.5269751 $ \\ \hline
Theor. value & $1.3841609326$ & $ - 6.1925057618568063655$ &
$34.526975286449930956$ \\ \hline
\end{tabular}

%

\caption{Solution to the Burgers equation with single-mode initial
condition: comparison of theoretical values with
6-stage and 13-stage asymptotic interpolation values for the 
first six coefficients in  Debye's solution \eqref{Debye1}.
For 13-stage asymptotic interpolation we also give some of the 
higher order coefficients.}
\label{t:9coeff}
\end{table*}


It is seen that the coefficients $C$, $\alpha$
and $\delta$ appearing in the leading term have a precision of at least
$10^{-10}$. The precision of the coefficients $\gamma_i$ for the 
subleading terms deteriorates with the order.

We now turn to the analysis using a 13-stage procedure. This allows a much
more precise determination of  the aforementionned coefficients and,
in principle, the determination of six
additional terms in the expansion \eqref{asgzero}.  After stage 6, the
next interpolation stage is stage 13. This is easily shown by
observing that the discrepancy between $G_n^{\rm (6)}$ and its
asymptotic value $3/\alpha =2$ is $O(1/n^4)$, because all the
lower-order terms have been stripped off by the first six 
transformations. More specifically, we have
\begin{eqnarray}
G_n^{\rm (6)} &=& 2 + \frac{c_1 }{n^4 } + \frac{c_2 }{n^5 } + \frac{c_3 }{n^
6 }
+ \frac{c_4 }{n^7 } + \frac{c_5 }{n^8 } + \frac{c_6 }{n^9 } +\ell_n\;,
\label{c1c6} \\
\ell_n &=&O\left(\frac{1}{n^{10} } \right)\; .
\label{rn10}
\end{eqnarray}
Stages 7--13 gives us the coefficients 
$c_1$,\ldots, $c_6$ and allow us to find the remainder $r_n$, as defined in 
\eqref{c1c6}. 
We found that   $r_n$, determined by this 13-stage procedure, falls to 
about $5\times 10^{-17}$ at the end of the range.  As shown in
Table~\ref{t:9coeff},  the
precision on the first  six coefficients in the asymptotic expansion
has improved very much and  is now of a few $10^{-17}$ for the
exponent $\alpha$.

\subsection{Further remarks on asymptotic interpolation}
\label{ss:joris-comments}


The method of asymptotic interpolation is still in the
development stage; improvements and new features are thus
to be expected. Some are already suggested in the initial
publication \cite{jorasint}. One rather straightforward extension is
from real to complex data. For rapidly growing data, one can use 
logarithms instead of ratios. In Ref.~\cite{jorasint} it is recommended
to take ratios or logarithms as often as possible and to define ``slow
growth'' as slower than, say, $n^{5/2}$. This helps
in identifying the functional form of the transseries expansion. Once
this is known, we  found that the values of the coefficients can
be generated more efficiently by using a rather broad definition
of slow growth, namely well-identifiable polynomial behavior.

\begin{figure*}
 \iffigs 
 \centerline{%
 \includegraphics[scale=1.0]{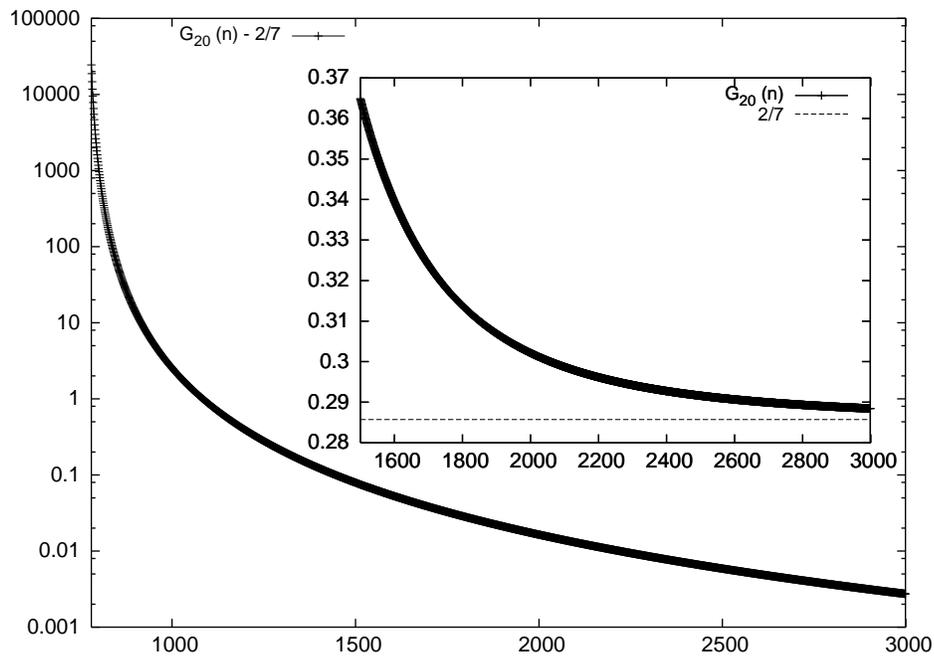}%
 }
 \else\drawing 65 10 {soc lin-log 2/1}
 \fi 
 
\caption{Interpolation stage $20$ has the asymptotic value $2/7$ as
shown in the inset. The main figure shows the discrepancy  $G_{20}(n) -2/7$.} 
\label{f:stage20}
\end{figure*}

A very important issue is to determine how many stages are feasible
with a given resolution $N$ and a given precision.  We have found that
the successive interpolation stages, at which the data are
asymptotically flat, have this flat regime preceded by longer
and longer transients. To make this more concrete we have investigated
how far it is necessary to go to be within five per cent of the
asymptotic value for the Burgers single-mode
problem.  For stage 6 the asymptotic value is 2 and the data are
within
five per cent everywhere. For stage 13, the asymptotic value is about 
$0.33836513$ 
and less than five per cent discrepancy  holds for $n>25$. The next
interpolation stage has number 20. The asymptotic value is $2/7$ 
but less than five per discrepancy  holds only beyond $n=1620$. In 
Fig.~\ref{f:stage20} we have represented  the data at stage 20.
It is seen that the discrepancy is enormous until we reach well
beyond $n=1000$. Obviously, if $N$ is 
not large enough the stripping of subleading terms performed by the
successive stages must be stopped \textit{however high the precision
of the data may be}.   A related issue is discussed at the end of
Section~7 of Ref.~\cite{jorasint}.

Since rounding errors increase with the stage
number, a certain balance must be kept between resolution and
precision. To investigate this quantitatively on the Burgers single-mode
problem, we have artificially
degraded our precision by adding random noise of various strengths.
It appears that we need at least 16 significant digits at stage 6 and 
27--35 significant digits at stage 13.

If we are only interested in obtaining an accurate determination of 
a few terms in the expansion
\eqref{asgzero}, we may be able to retrieve them using asymptotic
interpolation stopped at the sixth stage and continuing with a
different strategy. Indeed we observe that, at the sixth stage, the
data given by \eqref{rn10} have  the form 
\begin{equation}
G_n^{\rm (6)} = s + r_n\; ,
\label{srn}
\end{equation}
where the remainder $r_n$ decays to zero as $n\to\infty$. This is a
well studied situation in the theory of convergence acceleration by
sequence transformations, whose goal is to replace the sequence
\eqref{srn} by a transformed sequence having the same limit $s$ but a
much faster decaying remainder (see, e.g.,
Refs.~\cite{weniger,brez}). A simple and very popular 
acceleration method
appropriate for \eqref{rn10} is Wynn's rho-algorithm \cite{wynnrho},
although more sophisticated methods are known \cite{weniger}. In our case
it gives the correct value $2$ with a 20-digits precision.  This is
even better than the 13-stage asymptotic interpolation. Note that the choice of
a particular convergence acceleration method depends crucially on the
functional form of the remainder. With asymptotic interpolation this
form can be determined rather than having to be assumed.  Here 
a caveat is in order: if the data are not sufficiently asymptotic the
mixed procedure just described will not work, for example because the
remainder has not yet settled down to algebraic decrease. A situation of
this type seems to be present in the work 
on short-time asymptotics discussed in Ref.~\cite{paulsetal}: 
the rho-algorithm does not improve the quality of the scaling
exponent controlling the divergence of the vorticity at the singular
manifold  and much higher resolutions are probably needed for that problem.

E.J.~Weniger (private communication) has pointed out that asymptotic
interpolation and sequence transformations have technical features in
common. In asymptotic interpolation one tries to annihilate leading
terms in the asymptotic expansion, whereas in sequence transformation 
one tries to shrink the remainder by  annihilating its largest
contributions, but the transformations used in both instance are often the
same, for example, finite difference operators.

A powerful method of asymptotic series analysis, widely applied in statistical
physics, is the method of differential approximants, which can be viewed as a
generalization of the Dlog Pad\'e method \cite{guttmann}.  In particular it
has been used to analyze self-avoiding walks (SAWs) and polygons (SAPs).  We
have applied the asymptotic interpolation method to data available at
\texttt{http://www.ms.unimelb.edu.au/\,$\tilde{}$\,iwan/polygons/Pol\\ygons\_ser.html}. The
goal was to see how well we can reproduce the asymptotics of the number of
self-avoiding polygons with $2n$ steps on square and honeycomb lattices
\cite{jensgutt,jensen1,jensen2}. When analyzing the square lattice data for
the largest available range, that is $n$ up to 55, we found that asymptotic
interpolation gives the value of the critical point correct to 9 decimal
places, whereas differential approximants give about 3 additional 
digits. We observe that (i) the actual implementation with asymptotic
interpolation is somewhat simpler and (ii) asymptotic interpolation is not
limited to problems which can be well approximated by solutions of
low-order linear differential equations.

The method of asymptotic interpolation is, in our opinion, very useful
but is of course not the \textit{panacea}. One disease it cannot
directly cure is the presence of sinusoidal oscillations. For example
if the analytic function $f(Z)$ has  two complex conjugate
singularities at $R \ue ^{\pm \ui \phi_\star}$
on its circle of convergence, large-order  Taylor coefficients
will present a sinusoidal oscillation with a wavelength proportional
to $\phi_\star$. After any number of stages, this oscillation is still
present and the data cannot be interpolated by a constant. 
As we shall see now, a Borel transformation takes care of this
problem and can also bring hidden singularities to the foreground.

\section{P\'olya's theorem}
\label{s:polya}

Here we just want to give the reader
a good feeling of what the theorem states and a heuristic derivation.
We begin with examples discussed in Section~32 of Ref.~\cite{polya29}.

Let $c= |c| \ue ^{-\ui \gamma}$ be a complex number and 
consider the function
\begin{equation}
F(\zeta) = \ue ^{c\,\zeta} = 1+ \frac{c\,\zeta}{1!}+\frac{c^2\zeta ^2}{2!}+\ldots\;,
\label{example1F}
\end{equation}
which corresponds to the choice $a_n =c^n$ in \eqref{defFzeta}. The
Borel--Laplace transform, given by \eqref{BLdef}, is 
\begin{equation}
f^{\rm BL}(Z) = \frac{1}{Z}+\frac{c}{Z^2} +\frac{c^2}{Z^3 }+\ldots =\frac{1}{Z-c}\;.
\label{example1f}
\end{equation}
It has a pole at $Z=c$, whereas $F(\zeta)$ is an entire function 
(analytic in the whole complex domain). We set $\zeta =r \ue ^{\ui
  \phi}$
and let $r \to \infty$, holding the direction $\phi$ fixed; the
modulus of $F(\zeta)= \ue ^{|c| r \ue^{\phi-\gamma}}$, in the direction
$\phi$, varies
exponentially at the rate $h(\phi) = |c| \cos (\phi-\gamma)$, called
the \textit{indicatrix} of $F$. We define $k(\phi)\equiv {\rm Re}\,
\left( c \,\ue ^ {-\ui \phi} \right)  = |c| \cos (\phi+\gamma) $ . This is the (signed) distance
of the origin to the line normal to the direction $\phi$  passing through the pole $c$ and is called
the \textit{supporting function} of the (single)
singularity. We observe that 
\begin{equation}
h(\phi) = k(-\phi)\;.
\label{polyarelation}
\end{equation}
This relation is the simplest instance of P\'olya's theorem.

\begin{figure}
 \iffigs 
 \centerline{%
 \includegraphics[scale=0.7]{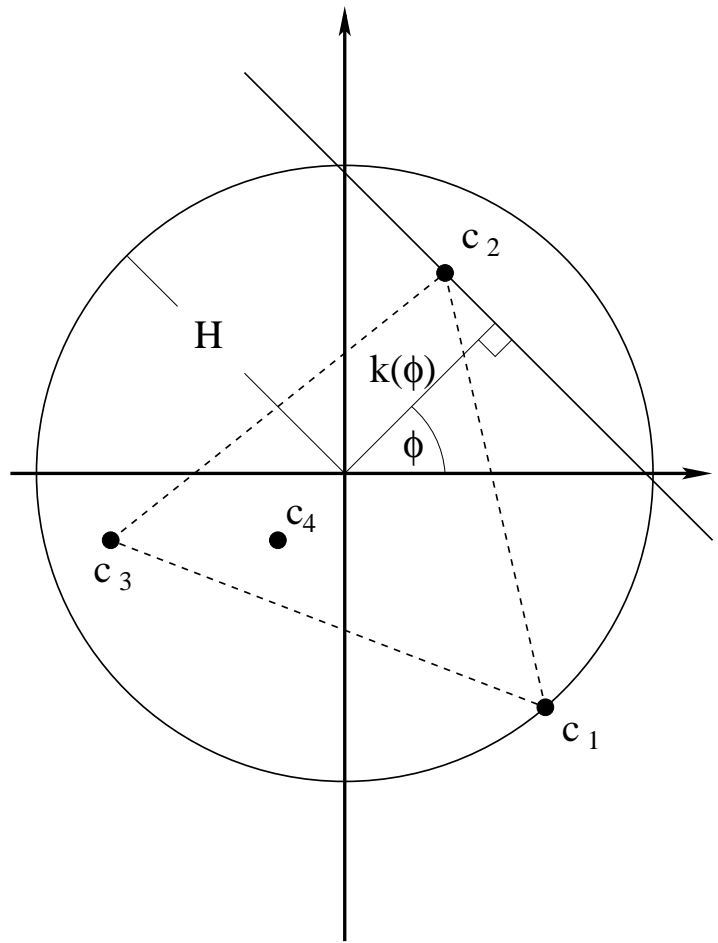}%
 }
 \else\drawing 65 10 {soc lin-log 2/1}
 \fi 
 \caption{Construction of the supporting function $k(\phi)$ of the set $K$ of 
singularities of $f^{\rm BL}(Z)$. The singularity $c_1$ is on the convergence circle; the convex hull 
of $K$ is defined by $c_1$, $c_2$ and $c_3$. The singularity $c_4$ is
inside the convex hull.} 
\label{f:convex}
\end{figure}

Next,  following again P\'olya's Section~32, we want to have $n$
distinct poles at the complex locations $c_1,\,c_2, \ldots ,\,
c_p$. For this we take complex linear combinations 
with non-vanishing coefficients
$C_1,\,C_2, \ldots ,\, C_p$: 
\begin{equation}
F(\zeta)= C_1 \ue ^{c_1\zeta} + C_2 \ue ^{c_2 \zeta}+\ldots 
+ C_p \ue ^{c_p \zeta}\;.
\label{example2F}
\end{equation}
The Borel--Laplace transform is 
\begin{equation}
f^{\rm BL}(Z)= \frac{C_1}{Z-c_1} + \frac{C_2}{Z-c_2}+\ldots+  \frac{C_p}{Z-c_p}\;.
\label{example2f}
\end{equation}
For any $\phi \in [0,2\pi]$, we define now the indicatrix and the
support function, a little more formally, as
\begin{equation}
h(\phi) \equiv 
\limsup_{r \to \infty} r^ {-1} \ln |F\left (r \ue ^{\ui \phi}\right)|\; ,
\label{defh}
\end{equation}
(in the present example, the $\limsup$ is  just an ordinary limit) and
\begin{equation}
k(\phi) \equiv \sup_{z\in K}(X \cos \phi +Y \sin \phi) =  \sup_{Z\in K} \left \{{\rm Re}\, \left(Z \ue ^{-\ui
  \phi}\right) \right \}\;,
\label{defk}
\end{equation}
where $Z= X +\ui Y$ and $K \equiv \{c_1,\, c_2,\ldots,\, c_p\}$ is the singular
set. Since
there is a finite number of singularities, the sup operation is just
the same as the maximum. We define a \textit{supporting line}
of $K$ as a line which has at least one point in common with $K$ and
such that all the points of  $K$ are in the same half space with
respect to the line. The intersection of all these half spaces is the
\textit{convex hull} of $K$. In the present case
this is just the smallest convex polygon containing all the poles. 
It  is readily seen
that $k(\phi)$ is the (signed) distance of the origin to the
supporting line normal to the direction $\phi$ (see
Fig.~\ref{f:convex}). 
The rate of growth of $F(\zeta)$ in the
direction $\phi$ is obviously that of the fastest growing of the $p$
exponentials in \eqref{example2F}, which is precisely $k(-\phi)$, so
that
P\'olya's relation \eqref{polyarelation} holds.

He proved a much more general theorem: \textit{Let $f^{\rm BL}(Z)$ be an
analytic function defined by the Taylor series \eqref{BLdef}
in powers of $1/Z$  which has  a finite
non-vanishing radius of convergence $H$, and let $K$ be  the smallest
convex compact set containing its singularities, then (i) the  Borel
transformed
series \eqref{defFzeta} defines an entire function of exponential
type, and (ii) the indicatrix $h(\phi)$ of $F(\zeta)$, defined by
\eqref{defh}, and the supporting function $k(\phi)$ of $K$, defined
by \eqref{defk}, are related by $h(\phi) =k(-\phi)$ and $H =
\sup_{\phi} h(\phi)$.}

(An entire function $F(\zeta)$ is said to be of exponential type if
its modulus is bounded by $A \ue ^{a |\zeta|}$, where $A$ and $a$ are
suitable positive constants.)

P\'olya's proof (not given here) makes use of the fact that $f^{\rm BL}$ and
$F$ are Laplace transformed of each other, specifically,
\begin{eqnarray}
&&f^{\rm BL}(Z) = \int_0^\infty F(\zeta) \ue ^{- Z \zeta}\,d\zeta\;, \label{Ftof}\\
&&F(\zeta) =\frac{1}{2\pi \ui} \int_{a -\ui \infty }^{a+\ui \infty}
  f^{\rm BL}(Z) \ue ^{Z \zeta}\, dZ\; ,
\label{ftoF}
\end{eqnarray}
where $a$ is any real number such that the singular set $K$ is entirely
contained in ${\rm Re}\, Z <a$.


Observe that no particular assumption is made regarding the type of the
singularities which can be isolated (e.g.\ poles or branch points) or
continuously distributed (natural boundary). Inside the circle of convergence
$|Z| =H$, the series \eqref{BLdef} is divergent. However if 
the whole circle is not a natural boundary,  the function $f^{\rm BL}(Z)$ can be
analytically continued to suitable $Z$'s inside this circle and the pair of 
integrals
\eqref{ftoF}-\eqref{Ftof} can be viewed as a way of resumming the
divergent series \eqref{BLdef}.

In applications it frequently happens that all the ``edge singularities'',
that is, those determining the border of the convex set $K$ are isolated. This
border is then piecewise linear, as in the case of $n$ poles discussed
above. The angular dependence of the supporting function is then given by
$k(\phi)= |c_j| \cos (\phi+\gamma_j) $ in the angular interval
$\phi_{j-1}<\phi<\phi_j$ for which the supporting line normal to $\phi$
touches $K$ at $c_j =|c_j| \ue ^{-\ui \gamma_j}$ (see Fig.~\ref{f:convex}). If $k(\phi)$
is known with high accuracy, then the positions of the edge singularities
$c_j$s can also be determined accurately.

\begin{figure}
 \iffigs 
 \centerline{%
 \includegraphics[scale=0.6]{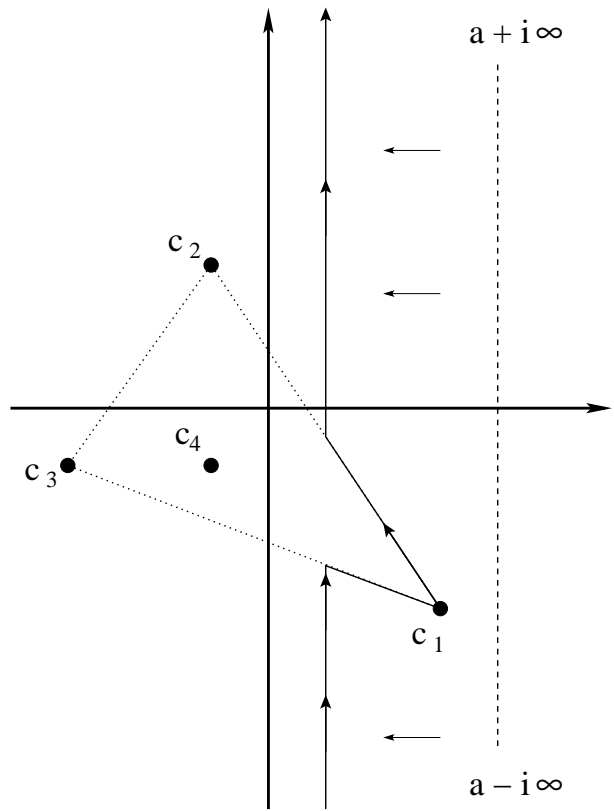}%
 }
 \else\drawing 65 10 {soc lin-log 2/1}
 \fi 
 
\caption{Contour of integration for computing the
inverse Laplace transform of $f^{\rm BL}(Z)$ (dashed line)
and its deformation to obtain the asymptotic contribution from the singularity 
at $c_1$ (continuous line with arrows).} 
\label{f:contour}
\end{figure}

Moreover we can then determine the type of an isolated  singularity at
$c_j$ by 
studying the asymptotic behavior of $F(\zeta)$ along rays $\zeta = r
\ue ^{\ui \phi}$ with large $r$, in a suitable angular interval. 
For example, 
let us assume that, near $c_j$
the function $f(z)$ has an algebraic singularity  and is to leading
order proportional to 
$(Z-c_j)^{\alpha -1}$, where the exponent
$\alpha$ is real and not a positive integer. (If $\alpha -1>0$, this behavior is assumed for a suitable first- or
higher-order increment of $f$.)
After shifting the contour of integration to follow the boundary of $K$
near $c_j$ (cf. Fig.~\ref{f:contour}), application of steepest descent
\cite{olver} to \eqref{ftoF} with $\zeta =
r\ue ^{\ui \phi}$ taken in the angular sector  $\phi_{j-1}<-\phi<\phi_j$ and $r\to\infty$
yields
\begin{eqnarray}
&&G(r)=|F(r\ue ^{\ui \phi})| = C r^{-\alpha}\ue ^{h(\phi) r}
[1+\varepsilon(r)],
\label{Fwithonesubleading}\\
&&h(\phi) = |c_j|\cos (\phi-\gamma_j)\;.
\label{hisolsingul}
\end{eqnarray}
Here  $C$ is a positive 
constant and $\varepsilon(r)$ tends to zero  for $r\to \infty$ at a rate which 
depends on what is assumed for subleading 
corrections to the $(z-c_j)^{\alpha -1}$ singular behavior. If we are
able to identify the
algebraic prefactor to the exponential in \eqref{Fwithonesubleading}, 
we can find the exponent $\alpha$ of the algebraic
singularity. 

Non-algebraic singularities can be handled similarly. For
example, if near $c_j$ the function $f(z)$ behaves as $\ue
^{1/(Z-c_j)}$,
application of steepest descent shows that instead of the algebraic
prefactor 
proportional to $r^{-\alpha}$ which appears in
\eqref{Fwithonesubleading}, we obtain an exponential prefactor 
proportional to $\ue ^{\pm 2\cos(\phi/2) \sqrt r}$. Furthermore, if all the
singularities on the convex hull of $K$ are isolated, then
\eqref{defk}
remains valid: the indicatrix is piecewise a cosine function.

How this is done in practice will be discussed in the next section.

\section{The Borel--P\'olya--Hoeven method}
\label{s:BPH}

As we have seen in Section~\ref{s:joris}, the asymptotic interpolation
method, applied to the Taylor coefficient of an analytic function
with a finite radius of convergence determined by a single isolated
singularity allows one to identify its  location and type. This is not the
case if there is more than one
singularity on the convergence circle. Furthermore  ``hidden''
singularities are not directly retrievable from the asymptotics of
Taylor coefficients. 

We can however take advantage of P\'olya's 
theorem (Section~\ref{s:polya}) to replace the analysis of  large-order
Taylor coefficients of an analytic function by the analysis of the
behavior of its  Borel transform at large distances in the
complex $\zeta$ plane along various rays.  This behavior can be found by
asymptotic interpolation, from which we can then construct the
convex
hull $K$ of the singularities  and  obtain their type when they are
isolated. This is the BPH strategy which we now describe in a 
more detailed way.

We start from a truncated  Taylor series in inverse powers of $Z$
\begin{equation}
f^{\rm BL}_{\rm T}(Z) = 
\sum_{n=0}^{N} \frac{a_n}{Z^{n+1}}\;,
\label{TBL}
\end{equation}
with $N$ terms, each of the coefficients being known with a precision $\varepsilon$.
We construct the associated truncated Borel series
\begin{equation}
F_{\rm T} (\zeta) \equiv \sum_{n=0}^{N} \frac{a_n}{n!} \zeta ^n\; .
\label{Tzeta}
\end{equation}

We choose a certain number of discrete angular directions characterized by 
their angle $\phi$. Along each ray $\zeta  = r \ue ^{\ui \phi}$,
we evaluate  the modulus of the Borel series at M points spaced by a
constant distance (mesh) $r_0$:
\begin{equation}
G_m \equiv |F_{\rm T} (m r_0\ue ^{\ui \phi})|,\qquad m =1,\,2,\ldots,\, M\;.
\label{defgm}
\end{equation}
Then we apply the asymptotic interpolation method of
Section~\ref{s:joris} to identify a large-$m$  leading-order behavior.
For example, for algebraic singularities we have 
\begin{equation}
G_m \simeq C(\phi) (m r_0) ^{-\alpha(\phi)} \ue ^ {h(\phi) mr_0}\;.
\label{gmasaympt}
\end{equation}
This gives us the constant $ C(\phi)$, the prefactor exponent $-\alpha(\phi)$
and the indicatrix $h(\phi)$ for the discrete set of directions. 
P\'olya's theorem then gives us the supporting function $k(\phi) = h(-\phi)$
of the set $K$ of singularities of the Taylor series. As we have seen in Section~\ref{s:polya}, if the singularities
on the convex hull of $K$ are isolated and are located at 
$|c_j|\ue ^{- \ui \gamma_j}$, then the supporting function is piecewise 
a cosine function,
given by $|c_j| \cos(\phi+\gamma_j)$. The exponent $\alpha$ gives us the
type of the singularity: a branch point (or a pole) of exponent
$\alpha -1$. 
Other types of singularities, for
example of the exponential type discussed near the end of
Section~\ref{s:BPH}, are handled similarly after identification of the
appropriate asymptotic behavior.

In practice, we have to choose the set of discrete directions, the 
mesh $r_0$ and the maximum number of points $M$ on each ray.
If we happen to know the  number $p$ of isolated singularities 
and, at least approximately, their positions we can  pinpoint the latter
by taking $2p$ suitable $\phi$ directions. This is however rarely the
case. We recommend taking a fairly large set of directions
(for example 500 uniformly spaced directions) in order to reduce the
risk of missing one or several of the cosine functions.
The natural choice for the mesh $r_0$ is $H^{-1}$ where  $H$ is the radius of 
convergence of the Taylor series. An approximate value is  $H_{\rm approx} = 
(1/n)\ln|a_n|$ for large $n$, which is roughly constant. For the determination
of the largest distance $r_{\rm max} =Mr_0$ we limit ourselves to the case where the function $F(r \ue ^{\ui \phi})$ 
grows at large distances, that is $h(\phi) >0$ 
(otherwise there are severe numerical problems).  $r_{\rm max}$ is then
determined by the condition that the last term $a_N \zeta ^N/N!$ of
the (truncated) Borel series \eqref{Tzeta} should introduce a relative
error in the determination of $F(r \ue ^{\ui \phi})$ which does
not exceed the precision $\varepsilon$ with which the Taylor
coefficients are known. A rough estimate for $|a_N|$ is $H^N$ and for
$|F(r_{\rm max} \ue ^{\ui \phi})$ is $\ue ^{r_{\rm max}H}$. Using the
Stirling formula, we find that, to leading order  $M \simeq N$ (the
dependence on $\varepsilon$ appears only in subleading logarithmic
corrections).

We mention that an improvement would be to replace a mere polynomial
truncation of the Borel series by a suitable resummation/acceleration method
for computing entire functions \cite{mueller}. This could be crucial for
determining negative indicatrix values, that is, when $F(\zeta)$ is
exponentially decreasing at large $\zeta$.

It is of interest to know how well we can separate two
discrete singularities. By P\'olya's theory, each singularity 
contributes an exponential term to the large-$r$ behavior
of the modulus of the Borel transform. If $r$ becomes sufficiently large
compared to the difference in the two e-folding rates, only one of the two 
singularities will be seen. By suitably changing the direction of the
ray in the $\zeta$-plane we can then focus separately on each
singularity.  The worst case for discrimination is when we  have two 
singularities which are at the same distance of the origin. Assuming that
this  common distance is comparable to the radius of convergence $H$ and
denoting by $\Delta$ the distance of the two singularities, we find that
the largest discrepancy in e-folding rate is roughly $\Delta ^2/H$. 
Denoting, as 
above, by $M$ the maximum number of point on a ray, we find that
good separation requires the separation parameter
$M\Delta ^2/H^2$ to be large. Since discrepancies are amplified
exponentially, a separation parameter of $10$ may suffice.

We shall not here discuss issues of algorithmic complexity, such as
the reduction of the number of operations to evaluate the
truncated Borel series. In applications the complexity of the 
numerical calculations
needed to accurately determine the Taylor coefficients will usually 
exceed very much what is needed for the BPH analysis.

\section{Testing BPH on the Burgers equation with multimode initial
conditions}
\label{s:burgers-multimode}

To test the BPH method we need a Taylor series having
either a pair of singularities on the convergence circle or 
``hidden singularities''. As in
Section~\ref{ss:joris-burgers}, this can be done using $2\pi$-periodic
solutions of the inviscid Burgers equation.
The 2-mode initial condition 
\begin{eqnarray}
&&u_0(a) = \lambda_1 \sin a + \lambda_2 \sin (2a)\;, \label{2mode}\\
&&\lambda_1 = -1/2 , \quad \lambda_2 = (1/16)(4-\sqrt(14)) + \epsilon\;,\nonumber\\
&&\epsilon = 1/150\nonumber
\end{eqnarray}
produces at $t=1$ a solution $u(1,z)$ having, in Eulerian coordinates,
singularities at
\begin{eqnarray}  
z_\star^\pm = &\pm& 0.1103542160016972443 \nonumber \\
&\pm& \ui\,
0.737097018253664793\;.
\label{2modezstar}
\end{eqnarray}
Henceforth we shall concentrate on the singularities of $u(1,z)$ in
the lower half plane, which are also the singularities of the function
$u^+(1,z)$, the sum of Fourier harmonics with $k>0$. Note that there are two
singularities with the same imaginary part
and opposite real parts (this is a consequence of the symmetry $a
\mapsto -a,\,\, u_0 \mapsto - u_0$ of the initial condition and of the
complex conjugate symmetry). When the Fourier series for $u^+(1,z)$
is transformed into a Taylor series in inverse powers of $Z$ by
setting $Z = \ue ^ {-\ui z}$, the $z$ singularities get mapped onto
two complex conjugate $Z$ singularities
\begin{equation}  
Z_\star^\pm =  0.4755903313336372343 \pm \ui \,0.0526974896343733942\;.
\label{2modeZstar}
\end{equation}

The 3-mode initial condition  
\begin{eqnarray}
&&u_0 (x) = \lambda _1 \sin x + \lambda _2 \sin 2x + \lambda _3  \sin 3x  
\label{3mode}\\
&&\lambda _1 = - \frac{1}{2},\quad  \lambda _2 = \frac{4-\sqrt{14} }{16} +
\frac{1}{50},\nonumber\\
&&\lambda _3 = - \frac{1}{100} 
\end{eqnarray}
produces at $t=1$ a solution $u(1,z)$ having, in Eulerian coordinates,
in the lower complex half plane singularities at
\begin{eqnarray}            
      z_{1\star} &=& -\ui\,   0.4608974136239120258
    \label{3modezstar1}\\
z_{2\star}^\pm &=&           \pm 0.8575677577466957833\nonumber \\
&~& -\ui\,               1.1175132271503113898\;.
\label{3modezstar2}
\end{eqnarray}

\begin{figure}
 \iffigs 
 \centerline{%
 \includegraphics[scale=0.4]{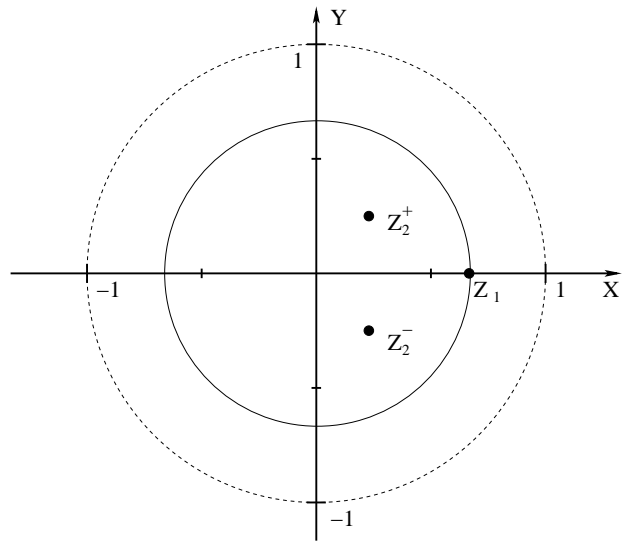}%
 }
 \else\drawing 65 10 {soc lin-log 2/1}
 \fi 
 
\caption{Positions in $Z$-plane of the singularities for the 3-mode
initial condition. Continuous line: circle of convergence;
dashed line: image of the real domain by the map $z \mapsto  \ue ^{-\ui z}$.} 
\label{f:3modesingul}
\end{figure}

The $z_{1\star}$ singularity is on the imaginary axis and is the
closest to the real domain. The other two are further away (hidden)
and symmetrically located with respect to the imaginary axis. The
corresponding
$Z$ singularities are 
\begin{eqnarray}
Z_{1\star} &=&     0.6307173770893952917\;,
\label{3modezstar1a}\\
Z_{2\star}^\pm &=& 0.2140094820693456182 \nonumber \\
&~& \!\!\!\pm \ui\,0.2473645913888956747 \;.
\label{3modeZstar2a}
\end{eqnarray}
are shown in Fig.~\ref{f:3modesingul}.

To apply the BPH method we generate the Fourier harmonics
with $k=1,\ldots,\, N =1000$ using the third Fourier--Lagrangian
representation \eqref{fl3}. The Lagrangian integrals are again 
calculated  with 80-digit precision 
and 120-digit working precision.  

The Borel transform is calculated for 20 values of $\phi$ between
$0$ and $\pi/2$ (a symmetry $\phi\to -\phi$ makes it unnecessary to
take $\phi <0 $ and for $\phi > \pi/2$ the indicatrix is negative).
For the mesh we take $r_0 =1$. The total number of points on each ray
is $M=500$.  Along each selected ray, application of six-stage
asymptotic interpolation gives us $C$, $\alpha$ and $h$. We can
check that the indicatrix is piecewise a cosine function as implied
by \eqref{defk}. Least square fits allow us to identify the parameters
of these cosine functions and thereby to find the locations of the
singularities.   We recover the known
values with an accuracy of about $10^{-6}$. We found that the accuracy
on ``hidden singularities'' is comparable to that on directly visible
ones. We also found that  $N=1000$ is not sufficiently asymptotic for a
13-stage analysis of the kind described
in Section~\ref{ss:joris-burgers}.

\section{Concluding remarks}
\label{s:conclusion}

One central theme  of this paper is the use of a Borel transform, in
conjunction with P\'olya's theorem, to reveal singularities not
directly accessible from the asymptotic behavior of the Taylor/Fourier
coefficients. A very useful property of
the Borel transform of a Taylor series (in inverse powers of $Z$) is
that its large-distance behavior encodes information not only about
those singularities of the Taylor series located on its convergence
circle, but also about other singularities ``hidden'' inside this
circle. Actually the Borel transform, followed by a Borel--Laplace
transformation is a way of performing analytic
continuation. Recovering hidden singularities from a Taylor series 
has important applications in a number of fields; many of
the known techniques have been reviewed by Guttmann
\cite{guttmann}. 

To the best of our knowledge P\'olya's theorem
has never been used as a numerical tool for identifying singularities. 
The theorem is of a very general nature and assumes nothing
about the nature of the singularities; this has the great advantage that we
do not have to distinguish between true and spurious singularities, as is
the case, for example, when using Pad\'e approximants and related methods.
The principal drawbacks are that (i) not all
hidden singularities are accessible, only those located on the convex
hull of the singular set,  (ii) pairs or clusters of  singularities 
situated to close to each other may not be easily distinguishable, and 
(iii) enough terms in the series must be
known to be able to actually obtain  the asymptotic behavior of the
Borel transform. When
hundreds to thousands of Taylor coefficients are known, alternative 
mathematically well-founded techniques may become competitive, for
example the old Weierstrass analytic continuation method; thanks
to recent algorithmic discoveries it can be performed quite 
efficiently \cite{Hoeven,joris-continuation}.

The other theme of this paper is the asymptotic interpolation method
of van der Hoeven which is here used both directly (when
Darboux's theorem is applicable) and indirectly by means of P\'olya's 
theorem. When a large number of Taylor/Fourier coefficients
are know with sufficient accuracy, asymptotic interpolation can
give truly remarkable results, providing us not only with very
accurate leading terms but also with several subleading corrections.
As we have seen in Section~\ref{s:joris}, there is usually a well-defined
relation between the number of subleading correction terms and the number
of stages of the procedure which can be achieved. The latter depends
crucially on the number of known coefficients and on their precision.
For example if the data have only double precision, it is unlikely
that more than six stages can be achieved. Asymptotic interpolation 
might than be viewed as an overkill compared to more standard
techniques,
but it is worth stressing that asymptotic interpolation is 
very easy to implement. 

Which kinds of problems are most likely to fall within the prongs of
full-strength Borel--P\'olya--Hoeven-type analysis?
This depends crucially on the computational complexity of the problem,
that is the dependence of CPU requirement and storage on the number of
coefficients $N$. As pointed out by Guttmann \cite{guttmann}, phase
transition problems formulated on a lattice require usually
enumerating diagrams and the number of these tends to grow
exponentially with order, while fluid dynamics problems generally have
only polynomial complexity. In connection with phase transitions our
BPH method is likely to be less precise than alternative methods
such as differential approximants, but it can usefully supplement them 
to ascertain that the singularities identified are not artefacts.

In fluid dynamics one outstanding problem is the issue of
finite-time blow-up for the three-dimensional incompressible Euler
flow with smooth initial data \cite{MB,blue}. For initial
data having simple trigonometric polynomial form, one can
determine numerically a number of coefficients of the Taylor expansion
in time of the enstrophy (integral of one-half the squared
vorticity). This was done for the Taylor--Green flow by Brachet et
al. \cite{BMONMF} (yielding 40 non-vanishing coefficients
calculated with quadruple working precision) and for the 
Kida--Pelz flow by Pelz and Gulak \cite{Pelz} (yielding 16
non-vanishing coefficients having at least 40-digit
precision). Because the number of coefficients is  rather small,
there is no consensus on what the results imply for blow-up. Such
calculations have a complexity $O(N^5)$ which can however be reduced
to $O(N^4)$ (up to logs) using the method of relaxed multiplication
\cite{Hoeven}. It is likely that a state-of-the-art calculation
for flows with simple trigonometric polynomial initial conditions can
give up to several hundred non-vanishing Taylor coefficients of the
enstrophy, with a working precision of several hundred digits. 
Another problem which can be tackled by series analysis is the
analytic structure of the two-dimensional incompressible vortex sheet 
(Kelvin--Helmholtz instability).  It is known that an initially analytic 
interface will develop a singularity in its shape after a finite time.
Moore has made a prediction regarding this singularity \cite{moore} 
which has been studied by various numerical techniques \cite{MBO,shelley}.
Again there is no consensus on the type of this singularity.

It is of course of interest to extend to several dimensions the BPH
method, here  presented only in the one-dimensional case. We observe that 
there exist multi-dimensional
generalizations of the Borel transform \cite{trutnev1,trutnev2} and that the
the asymptotic interpolation method can also in principle be extended
to several dimensions \cite{jorasint}. In several dimensions,
singularities are not point-like; they  reside on extended objects such
as analytic manifolds and can have a much more involved
structure than in one dimension. 
It is possible to partially reconstruct 
such objects using a variant of BPH. Furthermore, we note that 
P\'olya's theorem has been extended to several complex dimensions \cite{ronkin} 
(it is then referred to as the Ivanov--Stavski\u{\i} theorem). Information on
singularities can then in principle be obtained numerically in a way analogous 
to what has been done
in Section \ref{s:BPH}. This will be discussed elsewhere.

\vspace*{4mm}
\par\noindent {\bf Acknowledgments}

We are very much indebted to J.~van der Hoeven for having introduced
us to his asymptotic interpolation method and to E.J.~Weniger for his detailed
advice on extrapolation techniques. We have also benefited
very much from discussions with J.~Bec, M.~Blank, H.~Frisch, T.~Matsumoto,
D.~Mitra, R.~Pa\d{n}\d{d}it and A.K.~Tsikh.
\vspace{3mm}


\end{document}